\documentclass[aps,prb,preprint]{revtex4}

\begin{document}

\title{Experimentally probing the shape of extra dimensions}

\def\bar{\overline}

\author{Marc Sher}
\email{sher@physics.wm.edu}
\author{Kelly A. Sullivan}
\email{kelly@physics.ucsc.edu} 
\affiliation{Particle Theory Group, Department of Physics, College of
William and Mary, Williamsburg, Virginia 23187, USA}

\begin{abstract}
Theories with compact extra dimensions 
have become increasingly popular. In some theories the
standard model particles are confined to our four-dimensional world and
only gravity can propagate in the extra dimensions. In these models the
size of the extra dimensions can be as large as hundreds of microns. The 
functional dependence of the gravitational force on distance is well 
known at both large and small distances compared to the size of the 
extra dimensions; for two extra dimensions, for example, it varies 
from $1/r^{2}$ to $1/r^{4}$ as the distance decreases. However, the 
dependence for intermediate distances has not been fully calculated. We 
determine this dependence as a function of both the size of the extra 
dimensions and the possible angle between the extra dimensional unit 
vectors and show that high precision measurements of the 
gravitational force would make possible the determination of the shape
of the extra dimensions.
\end{abstract}

\maketitle

\section{Introduction}

Since the realization twenty years ago that string theories were 
promising candidates for a unified theory of all interactions, there 
has been extensive interest in theories with more than three 
spatial dimensions. In string theory consistency requires that 
there be nine spatial dimensions.\cite{stringreview} Because we only
observe three such dimensions, it is generally assumed that each of the
other six dimensions must be compact, curled into a very small size
$R$. For example, a cylinder has one large and one compact direction.  
For a small creature on the cylinder, the world looks two dimensional, 
while for a large creature, the cylinder looks one dimensional.  Thus 
one expects that at distance scales larger than $R$ (corresponding to energy scales
below $\hbar c/R$), the effects of these extra dimensions are unobservable. 
More precisely,
if $y$ is one of the compact dimensions, then $y$ is identified with 
$y+2\pi R$.  Thus, $f(y)=f(y+2\pi R)$ for any function.  By Fourier's 
theorem, any such function can be Fourier-expanded.  At distance scales larger than
$R$ (or energy scales smaller than $\hbar c/R$), only the zero mode, which
is independent of the extra dimensions, is relevant.  

The possibility of experimental detection of these extra dimensions 
has been considered negligible until recently, because it was
believed that the scale $R$ is far smaller than the scales that 
can be reached in colliders. As a particle's energy increases, its 
Compton wavelength shrinks, and at a high enough energy it will become
smaller than the particle's Schwarzschild radius (at which point
gravitational effects are important). Quantum gravitational 
effects will be critical at this energy scale, which is called the Planck 
energy. This energy scale  is
$\sqrt{\hbar c^{5}\over G}
\sim 10^{19}$\,GeV (corresponding to a Planck length of 
$10^{-33}$\,cm.). We thus expect the compactification scale $R$ to 
be of this order, which is far beyond the range of accelerators. 

The reason that the Planck energy is so 
high is that at macroscopic length scales gravity is much weaker 
than the other interactions and thus enormous energies must be reached 
before it becomes comparably strong.  A major problem in particle 
physics is trying to put into one single framework two widely  
different energy scales (the weak interaction scale of 
$100$\,GeV and the Planck scale).  This problem is called the hierarchy problem and is 
related to the relative weakness of gravity.

In the past few years, new models have appeared\cite{milli} in which the size of 
the extra dimensions can be much larger, as large as a millimeter! In these
models only gravitational interactions can propagate in the extra
dimensions; other particles and interactions are stuck in our 
three-dimensional space. In this way, only gravitational interactions at
distance scales less than a millimeter would be sensitive to these extra
dimensions. Having non-gravitational interactions stuck in our three-dimensional
isn't as ad hoc as it might appear --
string theories often have gravitons as closed strings and other particles
as open strings, and the open ends can be stuck on a three-dimensional
surface. Such large extra dimensions would have dramatic experimental 
implications. In higher dimensions, gravity would no longer be inverse 
square, but would be inverse-cube in four spatial dimensions, 
inverse-fourth in five, etc. This dependence follows from Gauss' law in
many dimensions. Thus, we would experimentally see gravity change from
being inverse-square at scales larger than $R$ to inverse-cubed at scales
smaller than $R$ (if there is one extra dimension).

We have explicitly assumed that Gauss' law will 
describe the gravitational interaction in many spatial dimensions. ÊThis
assumption follows naturally from string theories and is discussed in
Ref.~\onlinecite{milli}. Ê Gauss' law relates the integral of the force over
the surface of a volume to the amount of mass (or charge in the case of
electromagnetism) inside that volume. ÊÊ By the divergence theorem the
integral of the force over the surface can be converted into a divergence
of the force over the volume. The divergence of a force is the Laplacian of
the potential energy, which leads to Poisson's equation (giving $\nabla\cdot
\vec{E} = 4\pi \rho$ and Ê$\nabla^2 V = -4\pi\rho$ for electromagnetic
fields). ÊPoisson's equation follows directly from the Einstein field
equations in
$n$ dimensions, and thus Gauss' law follows from general relativity. ÊSee
Ref.~\onlinecite{milli} for details.

The major attractive feature of models in which there are large extra 
dimensions is that they can substantially decrease the Planck scale to accessible
energies. For example, suppose that there are six compact dimensions, all
with a size $R$ given by 20 fermis (corresponding to an energy scale of
10\,MeV). As the energy scale increases past 10\,MeV, gravity will change
from being 
$1/r^{2}$ to $1/r^{8}$. Its strength will then grow extremely 
rapidly at scales above 10\,MeV and will become strong at approximately 
a few\,TeV, which is within reach of currently planned accelerators. 
Thus the Planck scale will only be a few TeV, and we can expect 
future colliders to directly probe quantum gravity and string effects, 
produce black holes, etc. The traditional hierarchy problem is then 
resolved.

We can reverse the argument and determine the value of $R$ given 
that the requirement that the Planck scale be in the TeV range. This
requirement will depend only on the number of extra dimensions. For a
Planck scale of 
$1$\,TeV it is found\cite{milli}
that $R \sim 10^{32/d} 10^{-17}$\,cm., where $d$ is the number of 
extra dimensions.  For one extra 
dimension the value of $R$ turns out to be millions of kilometers, which
is clearly excluded. Thus, there cannot be just one extra dimension. For
two extra dimensions $R$ is a few hundred microns, which is
currently being experimentally probed.\cite{experiment,footn} For
three extra dimensions $R$ is tens of nanometers. Although this distance is 
accessible in some proposed experiments,\cite{nanoexperiment} reaching
force strengths comparable to the gravitational force seems far out of
reach. For more than three extra dimensions the value of $R$ is too small
to ever probe directly.

In this article we will focus on the case of two extra dimensions. With
more than one compact dimension, the shape is not completely constrained,
for example, with two extra dimensions, the ratio of the radii of the two
dimensions can be other than unity and the unit vectors need not
be orthogonal, leading to a shape angle. Thus, by precisely measuring the
force of gravity at small distances, 
we can probe the shape as well as the size of the extra dimensions.

\section{Calculational Methods and Gauss' Law}

Although we will focus on the case of two extra dimensions, we will 
first begin, as a toy model, with the  case of a single extra 
dimension, compactified with periodicity
$R$ ($R$ would be the circumference of a circle, not the radius). The
spatial variables are 
$x$, $y$, $z$, and $w$. A point mass in the system 
will appear as in Fig.~1, where $y$ and $z$ coordinates have been 
suppressed, and the periodicity in the compact dimension is 
apparent. It should be pointed out that we are working in a 
``brane-world'' scenario in which the known particles are confined 
to a four-dimensional brane and only gravity can propagate in the 
extra dimensions.

It is difficult to visualize four spatial dimensions. ÊTo 
simplify the visualization we consider a three-dimensional problem 
in which the masses in Fig.~1 are point charges and ask for the value of 
the electric field a distance $r$ from the central point charge. ÊFor $r << R$, 
only the central point charge is relevant, and the electric field is given 
by drawing a spherical Gaussian surface around the charge, resulting in a 
$1/r^2$ force. ÊAs $r$ increases, the other charges become important, and 
no simple Gaussian surface can be drawn; we must add the contributions 
explicitly, resulting in an infinite sum. ÊAs $r$ becomes much 
bigger than $R$, the point charges look like a line charge. ÊNow a 
cylindrical Gaussian surface can be used, resulting in a $1/r$ force law. Ê
An interesting exercise would be to write the sum explicitly, and show how 
it interpolates between the two different examples of Gauss' law (we will do
this interpolation for the higher dimensional case). ÊIn standard
electromagnetism texts the $1/r$ force law is derived by first considering
a finite length line and then increasing the length to infinity. ÊÊThe
method described here is an alternative
derivation that provides some insight into the $1/r$ behavior.

In four spatial dimensions and $r << R$, the gravitational (or 
electromagnetic) force law is $1/r^3$, which is found by encircling the 
point mass with a 4-sphere. ÊAs $r$ increases, the full sum must be used. Ê
At very large distances the points look like a line, and a Gaussian 
4-cylinder must be used. ÊNote that the surface area of a 3 cylinder 
(parallel to the axis) is given by the length of the cylinder times the 
circumference of the ``2-sphere" (that is, circle) endcaps, that is, $2\pi r
L$. Ê Thus the surface area of a 4-cylinder is the length times the area of
the 3-sphere endcaps, which is $4\pi r^2 L$. ÊWe now proceed to the 
calculation of the gravitational potential from the masses in Fig.~1 in 
4 spatial dimensions.

For very small distance scales, $r << R$, the gravitational field 
depends only on the mass at $w=0$. Gauss' law in $n$-dimensions\cite{landy}
is
\begin{equation}
\int\! F\cdot da = S_nG_nM_{\rm enc},\label{eq:1}
\end{equation}
where $S_{n}$ is the surface area of a unit $n$-sphere, $M_{\rm enc}$ is the 
mass enclosed in the Gaussian volume, and $G_{n}$ is 
the $n$-dimensional Newton's constant. Note that Newton's constant 
can be defined through this equation-- the constant $S_{n}$ is
only a convention. In three spatial dimensions 
the left-hand-side of Eq.~(\ref{eq:1}) gives $F 4\pi r^{2}$, and the 
right-hand-side gives $4\pi G_{3}M$. By equating these two terms, we find
the familiar Newton's law for the gravitational field. In higher 
dimensions, the surface area of a unit sphere is 
\begin{equation}
S_n = \frac{2 \pi ^{n/2}}{\Gamma(\frac{n}{2})},
\end{equation}	
where $\Gamma$ is the Gamma Function.  For our purposes, it is 
sufficient to know that $\Gamma(n)=(n-1)\Gamma(n-1), \Gamma(1)=1, 
\Gamma({1\over 2})=\sqrt{\pi}$.   We see that the gravitational field will 
be given by
$F= G_{n}M/r^{n-1}$. For one extra dimension, we see this reasoning gives
$F=G_{4}M/r^{3}$. Note that the dimensionality of $G_{4}$ and 
$G_{3}$ (the usual Newton's constant)
are different.

At very large distances, $r >> R$, the string of masses looks like a 
continuous line of uniform mass density, and we can 
use a cylindrical Gaussian surface to solve for the 
field.\cite{milli} Consider 
a 4-dimensional cylinder of side length $L$, with endcaps composed of 
three dimensional spheres of radius $r$. The mass enclosed by the cylinder
is $M {L\over R}$, $S_{4}=2\pi^{2}$, 
and the left-hand-side of Gauss' law is $F 4\pi r^{2} L$. 
Plugging these into Eq. (1),  the $L$ drops out (as usual in
cylindrical Gauss' law applications), and we find that 
\begin{equation}
F=G_{4}M{2\pi^{2}\over 4\pi}{1\over r^{2}R}.
\end{equation}
Thus at large distances, we recover the usual 
inverse square law. Comparing this expression with the conventional 
Newton's Law of gravity, we can find
the expression for $G_{4}$ in terms of $G_{3}$ (the usual gravitational
constant) and 
$R$. In $n$ spatial dimensions, we can write
\begin{equation}
G_{3} = {S_{n}\over 4\pi}{G_{n}\over V_{n-3}}
\end{equation}
where $V_{n-3}$ is the volume of the $(n-3)$-dimensional compactified
space. 

Thus, we find from Gauss' law that the gravitational field varies 
from an inverse-square law at large distances to an inverse ($n-1$) law 
at small distances, where $n$ is the number of spatial dimensions. 
This result is independent of the precise shape of the extra 
dimensions. To probe the shape we must look at intermediate 
distances. In this case, we must explicitly sum over the ``mirror 
masses'' and thus Gauss' law does not provide a useful way to calculate the
force.  For simplicity, we 
will calculate the gravitational potential at intermediate 
distances -- the field can be determined by differentiation. Again, we 
first consider a single extra dimension as in Fig.~1. 
In four spatial dimensions the gravitational potential at a 
distance $D$ from a point mass is 
$V_{4}=-G_{4}M/2D^2$ and so the general result for the 
potential a distance $r = \sqrt{x^{2}+y^{2}+z^{2}}$ from the mass $M$ is
\begin{equation}
V(r) = -{G_{4}M\over 2}\!\sum_{n=-\infty}^{\infty}{1\over r^{2}+n^{2}R^{2}}.
\label{eq:5}
\end{equation} 
In terms of the conventional Newton's constant $G_{3}$, Eq.~(\ref{eq:5}) 
becomes
\begin{equation}
V(r) = -{G_{3}M\over \pi R}\sum_{n=-\infty}^{\infty}{1\over 
n^{2}+\Delta^{2}},
\end{equation}
where $\Delta\equiv r/R$. For large $\Delta$, the sum is 
$\pi/\Delta$, leading to the conventional gravitational 
potential.

We plot $rV(r)$ as a function of $r$ in Fig.~2. As 
expected, the potential varies as $1/r$ for large $r$ and as 
$1/r^{2}$ for small $r$. Because one extra dimension is already excluded, 
we now consider the case of two extra dimensions must.

\section{Two Extra Dimensions}

With two extra dimensions the topology of the compactified space is 
no longer simple. Because the three dimensions of our universe are 
topologically flat, we will assume that the two compact dimensions are 
also flat. This assumption occurs naturally in string theory. Flat
compact dimensions are not unnatural. For example, the surface of a
cylinder is two-dimensional, and has one very large and one very small 
dimension. Yet it is topologically flat -- the angles of a triangle on 
a cylinder always sum to $180$ degrees. 

Technically, the space we 
are considering is 
the product of two circles, known as a 2-torus. It can have two 
circumferences, which we will denote as $R_{1}$ and $R_{2}$. In 
addition, there is no reason that the unit vectors in the two compact 
directions must be orthogonal.\cite{shape} Thus, the description of the image 
masses corresponding to Fig.~1 is given in Fig.~3. The 
three parameters that specify the shape of the extra dimensions are 
$R_{1}$, $R_{2}$, and $\theta$.

As before, we can look at a 3-dimensional electromagnetic analogy with the 
masses in Fig.~3 replaced by point charges. ÊAt very short distances, only
the central charge contributes, giving the $1/r^2$ force. Ê At intermediate
distances, we must calculate the full double sum. At very large distances,
a Gaussian pillbox will yield a constant field. Ê An interesting
exercise would be to show how the double sum explicitly results in
Coulomb's law at short distances and a constant field at large distances.

We will first consider the case in which the unit vectors are 
orthogonal. The potential is given by
\begin{equation}
V=-{G_{5}M\over 3}\sum_{m=-\infty}^{\infty} 
\sum_{n=-\infty}^{\infty}{1\over 
(r^{2}+n^{2}R_{1}^{2}+m^{2}R_{2}^{2})^{3/2}},
\end{equation} 
where without loss of generality we can choose $R_{2}\geq R_{1}$; 
$G_{5}$ is Newton's constant in five spatial dimensions.

The result is plotted in Fig.~4 for $R_{2}=R_{1}$, $R_{2}=3R_{1}$, and 
$R_{2}=10R_{1}$. Note that at large distances, the 
standard $1/r$ potential emerges as expected. For small distances, 
we can clearly see the $1/r^{3}$ dependence of the potential. It is important to note that a measurement of the 
strength
of the gravitational potential at very small distances would enable 
us to distinguish between the three possibilities. One expects these 
possibilities to be distinguishable,  because the strength of the potential depends on the 
five-dimensional Newton's constant, which depends on the volume of the 
extra dimensions (see Eq.~(4)). However, the volume of the 
extra dimensions does not determine the shape -- we could not 
distinguish between $R_{1}=R_{2}=\sqrt{10} R_{0}$ and 
$R_{2}=10R_{1}=10R_{0}$, where $R_{0}$ is an unknown scale. Thus, 
to explore the shape of the extra dimensions, the full position 
dependence of the potential must be explored.

In Fig.~5 we have rescaled $R_{1}$ and $R_{2}$ so that the volume 
of the extra dimensions is the same in each of the three cases. We 
can now determine the precision necessary to distinguish the three 
cases. From Fig.~5 we can see that the value of the potential 
varies as $R_{2}$ varies from $R_{1}$ to $10R_{1}$ by a little over 
a factor of two. A 30\% measurement of the potential would enable 
us to distinguish between $R_{2}=R_{1}$ and $R_{2}=3R_{1}$. Such 
precision seems experimentally feasible, once a deviation 
from the inverse-square law is found.

 The full expression for
the potential at intermediate distances has not previously  been plotted. 
ÊIn many other
papers (see Refs.~\onlinecite{milli}, \onlinecite{floratos}, and
\onlinecite{cern}), the leading order deviation as 
$r$ is  decreased is given. This deviation is of
the form $V=V_0[1 + \alpha \exp(-r/R)]$, and many experimenters 
make plots of the allowed region in the $\alpha$-$R$ plane. ÊBut a 
measurement of a nonzero value of $\alpha$ would only establish a deviation --- 
the expression breaks
down for slightly smaller values of $r$ --- and the full expression would
then be necessary.

The other possibility is that the unit vectors in the extra dimensions 
are not orthogonal. Here, we assume $R_{1}=R_{2}$ for simplicity and 
vary $\theta$. If we again rescale the values of $R$ so that the volumes
are identical, we find the results in Fig.~6. We see that it is 
essentially impossible to distinguish between $\theta=\pi/2$ and 
$\theta=\pi/3$, but that a factor of two measurement of the potential 
would distinguish between these cases and $\theta=\pi/36$.

\section{Discussion}

Experiments probing the nature of gravity at short distances are now 
underway (see Ref. ~\onlinecite{experiment} for a review).  Measurements at the nanometer-scale are
possible. Should a deviation from the inverse-square law be discovered,
there will be a huge effort to measure this deviation precisely. In large
extra dimension models, the nature of the deviation is well known at very 
short distances, and the leading order term in the deviation is also 
well known. In this paper we have discussed the deviation in the 
intermediate region and have shown that experimentally constraining 
the shape as well as the volume of the extra dimensions is 
feasible. For two extra dimensions, in which the 
expected scale is in the tens to hundreds of microns region, a 
measurement of $30\% $ accuracy in the potential would enable us to 
distinguish a difference of a factor of three in the size of the extra 
dimensions, and shape angles smaller than a few degrees would also be 
measurable.

There are several possible avenues for future study. We could 
consider three extra dimensions, which would be aesthetically 
appealing because the nine spatial dimensions of string theory could 
be grouped into three large, three intermediate, and three small 
dimensions. In this case there would be three different sizes as well as 
three different shape angles. As noted, it is 
unlikely that this case is measurable experimentally.

Another possibility is that the extra dimensions are not flat. Here 
we have assumed that they are flat and form a 2-torus, the simplest case.
However, it could be that the extra dimensions are curved. For example,
they could have the topology of the surface of a sphere.\cite{cern} In this
case the simple description of Fig.~3 would not be accurate (because one
can travel in any direction and return to one's starting point). An
expansion in terms of Kaluza-Klein modes on a sphere would be appropriate
in this case. This expansion was done in Ref.~\onlinecite{cern}, where the 
leading order deviation from Newton's law is presented.

\begin{acknowledgments}
We are grateful to Josh Erlich for many useful suggestions. This work was
supported by the National Science Foundation under grant PHY-0243400.
\end{acknowledgments}

\newpage
\section*{Figure Captions}

\begin{figure}[h]
\caption{A point mass in a space with one compact dimension $w$ and 
one noncompact dimension $x$.}
\end{figure}

\begin{figure}[h]
\caption{The gravitational potential times $r$ is plotted as a 
function of $r$ for one extra dimension. Note that for 
large distances, the potential varies as $1/r$, and for short 
distances as $1/r^{2}$. The straight line has slope $-1$.}
\end{figure}

\begin{figure}[h]
\caption{A point mass located in a space of two compact dimensions of 
circumferences $R_{1}$ and $R_{2}$, with an angle $\theta$ between 
the unit vectors.}
\end{figure}

\begin{figure}[h]
\caption{The gravitational potential, times $r$ as a function of 
distance for various values of $R_{2}/R_{1}$. We have fixed $R_{1}$ 
and thus the volume of the compact space will vary in each case.}
\end{figure}

\begin{figure}[h]
\caption{The gravitational potential times $r$ as a function of 
distance, where the value of $R_{1}$ has been rescaled so that the 
volume of the compact dimensions are unchanged as $R_{2}$ is varied.}
\end{figure}

\begin{figure}[h]
\caption{The gravitational potential times $r$ for $R_{1}=R_{2}$ 
for various values of the shape angle. The results have been 
scaled so that the volumes are identical.}
\end{figure}


\begin{thebibliography}{99}
\bibitem{footn} For a Planck scale of 1\,TeV, the value of $R$ is a
millimeter; for a Planck scale of 10\,TeV, it is ten microns.
\end{thebibliography}
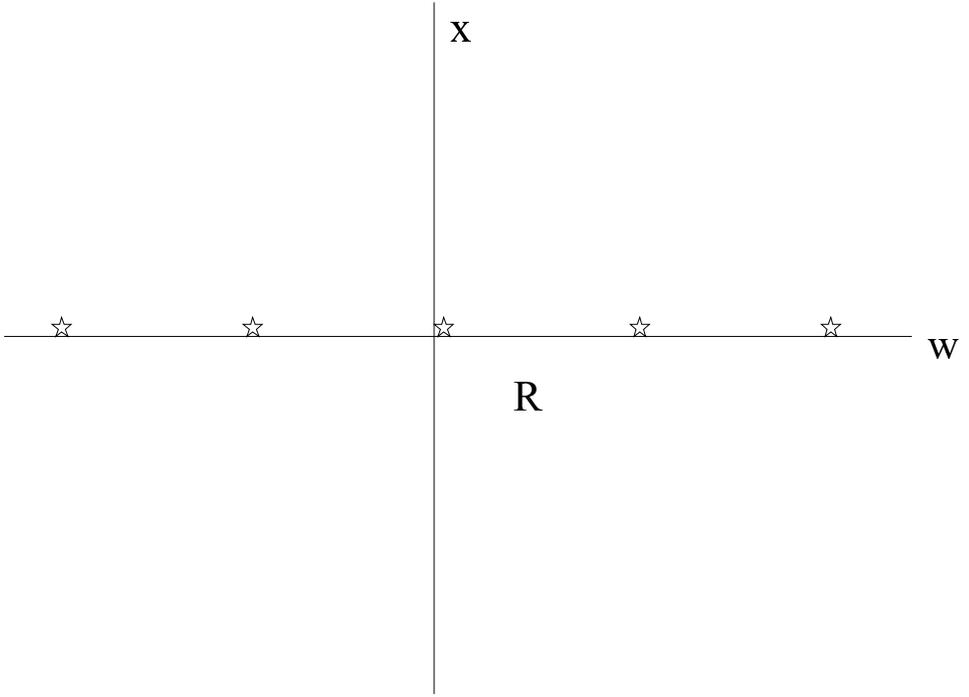
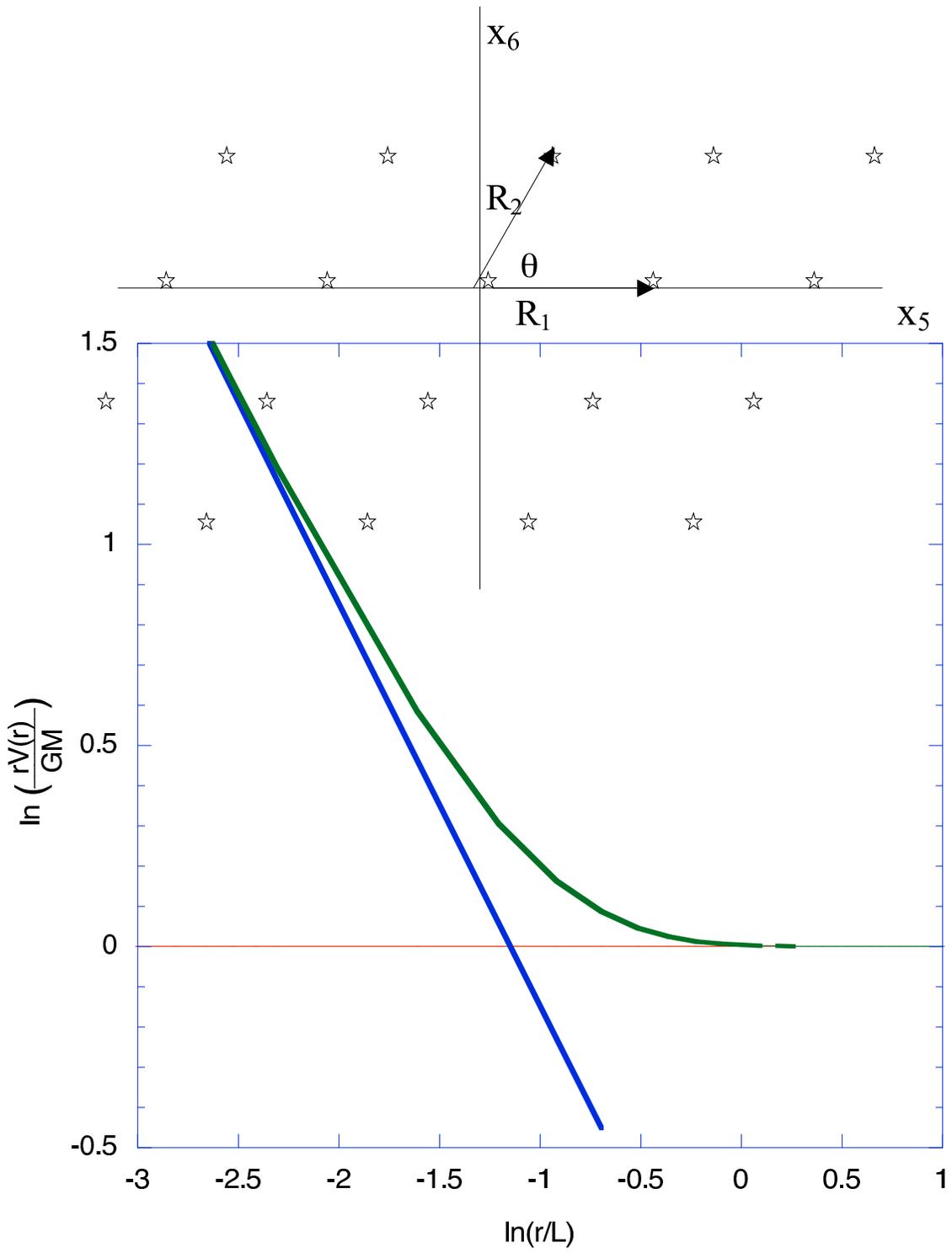
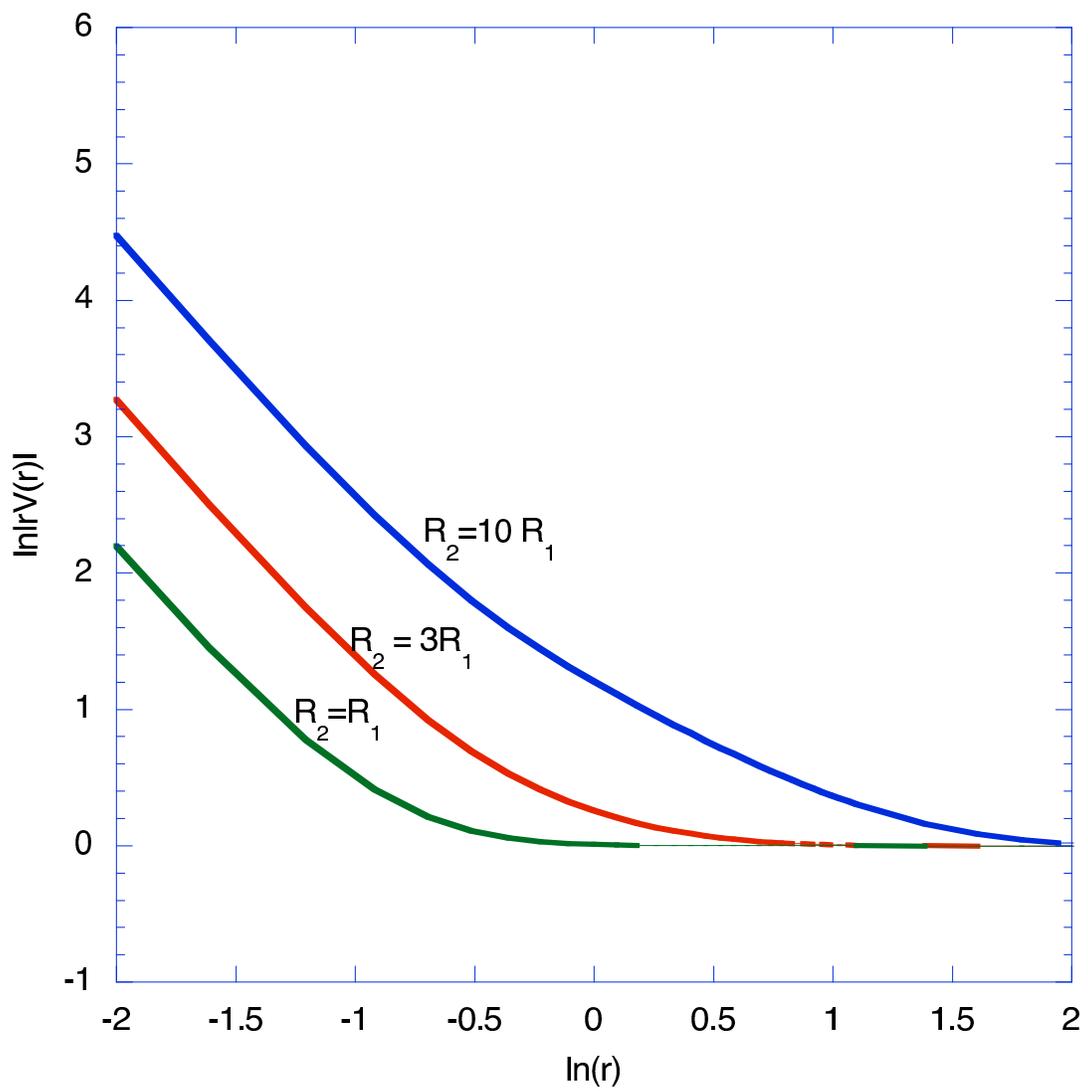
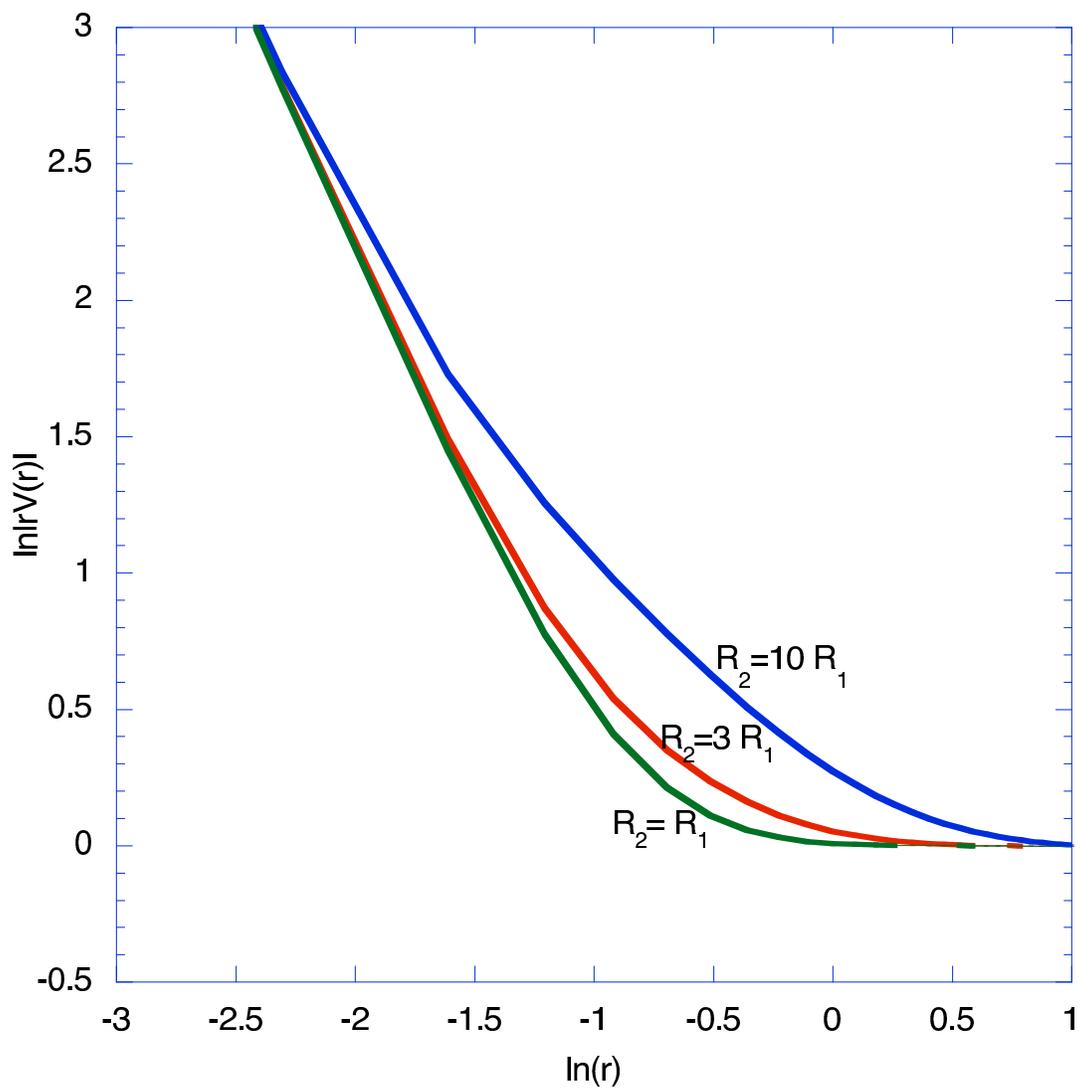
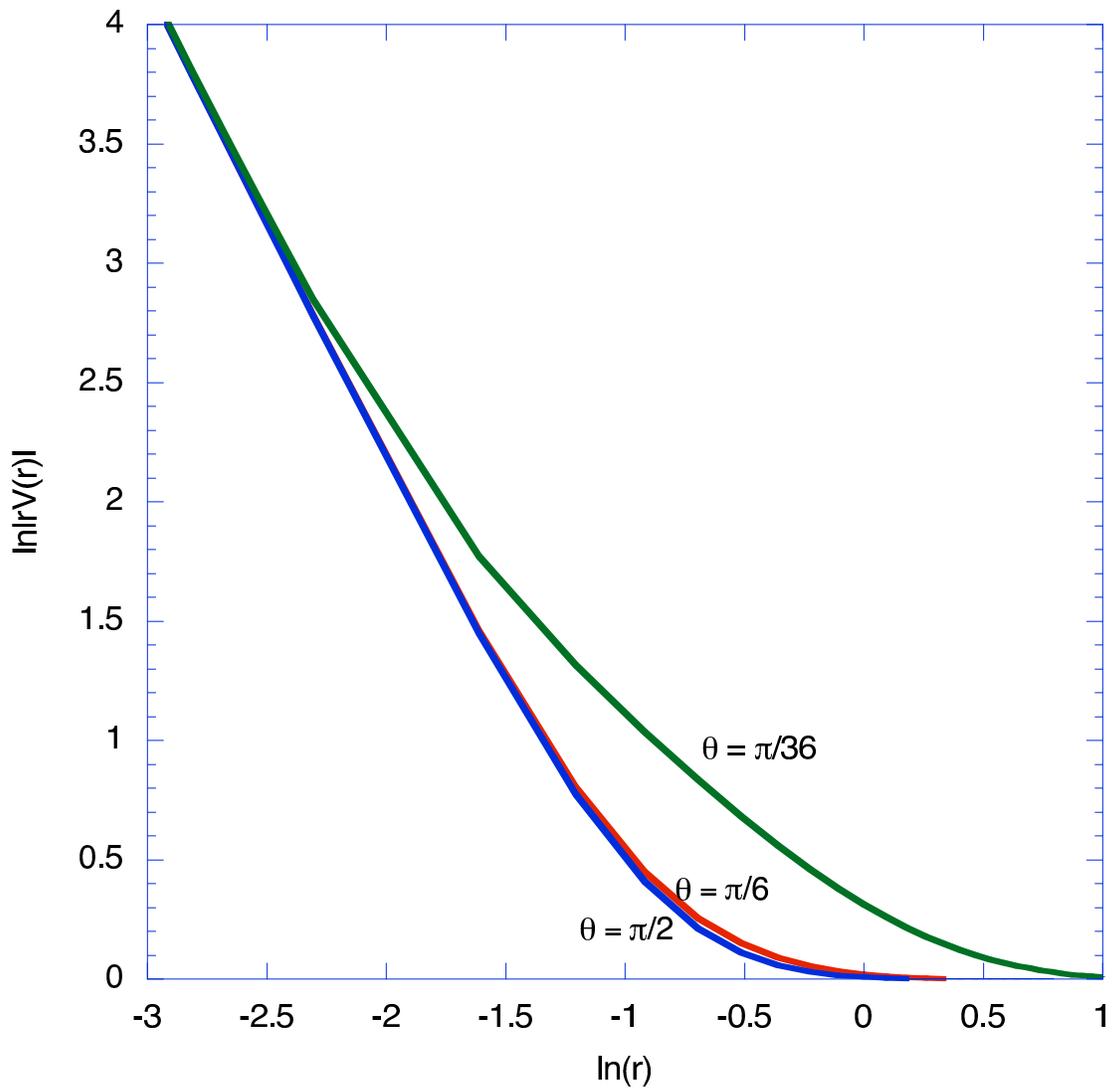
\end{document}